\documentclass[preprint,12pt]{elsarticle}

\usepackage{amssymb}
\usepackage{fullpage}
\usepackage{hyperref}
\usepackage{xcolor}
\journal{NIM B}

\begin{document}

\begin{frontmatter}

%% Title, authors and addresses

%% use the tnoteref command within \title for footnotes;
%% use the tnotetext command for theassociated footnote;
%% use the fnref command within \author or \address for footnotes;
%% use the fntext command for theassociated footnote;
%% use the corref command within \author for corresponding author footnotes;
%% use the cortext command for theassociated footnote;
%% use the ead command for the email address,
%% and the form \ead[url] for the home page:
%% \title{Title\tnoteref{label1}}
%% \tnotetext[label1]{}
%% \author{Name\corref{cor1}\fnref{label2}}
%% \ead{email address}
%% \ead[url]{home page}
%% \fntext[label2]{}
%% \cortext[cor1]{}
%% \affiliation{organization={},
%%             addressline={},
%%             city={},
%%             postcode={},
%%             state={},
%%             country={}}
%% \fntext[label3]{}

\title{Voltage scanning and technical upgrades at the Collinear Resonance Ionization Spectroscopy experiment}

%% use optional labels to link authors explicitly to addresses:
%% \author[label1,label2]{}
%% \affiliation[label1]{organization={},
%%             addressline={},
%%             city={},
%%             postcode={},
%%             state={},
%%             country={}}
%%
%% \affiliation[label2]{organization={},
%%             addressline={},
%%             city={},
%%             postcode={},
%%             state={},
%%             country={}}

\author[inst1,inst2]{Michail Athanasakis-Kaklamanakis\corref{cor1},\fnref{label1}}
\author[inst3]{Jordan R. Reilly\corref{cor2}\fnref{label1}}
\author[inst1]{\'{A}gota Koszor\'{u}s}
\author[inst4,inst11]{Shane G. Wilkins}
\author[inst2]{Louis Lalanne}
\author[inst2]{Sarina Geldhof\fnref{label2}}
\author[inst5]{Miranda Nichols}
\author[inst6]{Quanjun Wang}
\author[inst2]{Bram van den Borne}
\author[inst3]{David Chorlton}
\author[inst2]{Thomas E. Cocolios}
\author[inst3,inst7]{Kieran T. Flanagan}
\author[inst4]{Ronald F. Garcia Ruiz}
\author[inst2]{Ruben de Groote}
\author[inst5]{Dag Hanstorp}
\author[inst2]{Gerda Neyens}
\author[inst3]{Andrew J. Smith}
\author[inst4]{Adam R. Vernon}
\author[inst10]{Xiaofei F. Yang}

\cortext[cor1]{Corresponding author: m.athkak@cern.ch}
\cortext[cor2]{Corresponding author: jordan.reilly@cern.ch}
\fntext[label1]{These authors contributed equally to the work.}
\fntext[label2]{Current address: GANIL, CEA/DRF-CNRS/IN2P3, B.P. 55027, 14076 Caen, France}

\affiliation[inst1]{organization={Experimental Physics Department, CERN},%Department and Organization
            postcode={CH-1211 Geneva 23},
            country={Switzerland}}
\affiliation[inst2]{organization={Instituut voor Kern- en Stralingsfysica, KU Leuven},%Department and Organization
            postcode={B-3001 Leuven}, 
            country={Belgium}}
\affiliation[inst3]{organization={School of Physics and Astronomy, The University of Manchester},%Department and Organization
            postcode={Manchester M13 9PL}, 
            country={United Kingdom}}
\affiliation[inst4]{organization={Department of Physics, Massachusetts Institute of Technology},%Department and Organization
            city={Cambridge, MA},
            postcode={02139}, 
            country={USA}}
\affiliation[inst11]{organization={Laboratory for Nuclear Science, Massachusetts Institute of Technology},%Department and Organization
            city={Cambridge, MA},
            postcode={02139}, 
            country={USA}}
\affiliation[inst5]{organization={Department of Physics, University of Gothenburg},%Department and Organization
            postcode={SE 412 96 Gothenburg}, 
            country={Sweden}}
\affiliation[inst6]{organization={School of Nuclear Science and Technology, Lanzhou University},%Department and Organization
            postcode={Lanzhou 730000}, 
            country={China}}
\affiliation[inst7]{organization={Photon Science Institute, The University of Manchester},%Department and Organization
            postcode={Manchester M13 9PY}, 
            country={United Kingdom}}
\affiliation[inst10]{organization={School of Physics and State Key Laboratory of Nuclear Physics and Technology, Peking University},%Department and Organization
            postcode={Beijing 100971}, 
            country={China}}

\begin{abstract}
To optimize the performance of the Collinear Resonance Ionization Spectroscopy (CRIS) experiment at CERN-ISOLDE, technical upgrades are continuously introduced, aiming to enhance its sensitivity, precision, stability, and efficiency. Recently, a voltage-scanning setup was developed and commissioned at CRIS, which improved the scanning speed by a factor of three as compared to the current laser-frequency scanning approach. This leads to faster measurements of the hyperfine structure for systems with high yields (more than a few thousand ions per second).
% which reduced the required measurement time on Ag and Al by more than a factor of 3 with no loss in efficiency or resolution. 
Additionally, several beamline sections have been redesigned and manufactured, including a new field-ionization unit, a sharper electrostatic bend, and improved ion optics. The beamline upgrades are expected to yield an improvement of at least a factor of 5 in the signal-to-noise ratio by suppressing the non-resonant laser ions and providing time-of-flight separation between the resonant ions and the collisional background. Overall, the presented developments will further improve the selectivity, sensitivity, and efficiency of the CRIS technique.
\end{abstract}

%%Graphical abstract
% \begin{graphicalabstract}
% \includegraphics{grabs}
% \end{graphicalabstract}

%%Research highlights
% \begin{highlights}
% \item Research highlight 1
% \item Research highlight 2
% \end{highlights}

\begin{keyword}
%% keywords here, in the form: keyword \sep keyword
 CRIS \sep collinear resonance ionization spectroscopy \sep laser spectroscopy \sep voltage scanning \sep ISOLDE
%% PACS codes here, in the form: \PACS code \sep code
% \PACS 0000 \sep 1111
%% MSC codes here, in the form: \MSC code \sep code
%% or \MSC[2008] code \sep code (2000 is the default)
% \MSC 0000 \sep 1111
\end{keyword}
\end{frontmatter}

\section{Introduction} \label{sec:intro}
Laser spectroscopy of radioactive atoms and ions has been a cornerstone of ground- and long-lived isomeric-state nuclear studies for several decades~\cite{Campbell2016,Yang2022}. By measuring the hyperfine structure and isotope shifts of an electronic transition across multiple isotopes of the same element, the nuclear electromagnetic moments, nuclear spin, and changes in mean-squared nuclear charge radii can be measured. Systematic studies of these observables can then be used to study the strong nuclear interaction and the emergence of nuclear phenomena as a function of the nucleon numbers.

Traditionally, two approaches to laser spectroscopy have been most widely used on radioactive species: in-source resonance ionization spectroscopy (RIS)~\cite{Marsh2013} and fluorescence-detection collinear laser spectroscopy (CLS) of fast beams~\cite{Neugart2017}. 
As a merger of the two approaches, the collinear resonance ionization spectroscopy (CRIS) experiment at ISOLDE combines the high sensitivity of RIS with the high resolution of CLS. CRIS has thus successfully studied species with a production rate down to 20 nuclei per second~\cite{deGroote2020b}, while also reaching a resolution of 20(1)~MHz in the Fr chain~\cite{deGroote2015}. 
Recently, CRIS also ventured into the spectroscopy of radioactive molecules that are of interest for the study of fundamental symmetries~\cite{GarciaRuiz2020,Udrescu2021}. Implementations of the CRIS technique are currently also under construction and commissioning at the IGISOL laboratory of the University of Jyv\"{a}skyl\"{a}, Finland and at the Facility for Rare Isotope Beams (FRIB), USA.

In this article, recent technical upgrades at the CRIS experiment are described. The commissioning of a voltage-scanning apparatus has demonstrated that narrowband spectroscopy at CRIS can be now performed at a 3-6 times faster rate, for cases that are not statistically limited by a low production yield (typically with yields above a few thousand ions per second). Moreover, upgrades on several sections of the CRIS beamline are expected to significantly improve the signal-to-noise ratio of the technique.

\section{CRIS technique} \label{sec:technique}
In the CRIS technique, the fast, isobarically pure, cooled and bunched short-lived ions delivered by ISOLDE are firstly neutralized in a charge-exchange cell (CEC) filled with alkali vapor. The residual ions are deflected away from the beam, while the neutral atoms are transferred to the interaction region (IR). In the IR, the atoms are spatially and temporally overlapped with a series of laser pulses that step-wise resonantly excite an electron from the ground state (or a metastable state) to above the ionization potential. To avoid collisional (non-resonant) ionisation, the IR and the remaining part of the beamline are kept at a vacuum below $5\times10^{-9}$~mbar. The ions are then deflected away from the residual neutral species (e.g. neutralized isobars) and onto a MagneToF single-ion detector, and the signal is then processed with a time digitizer with sub-ns resolution.

In a typical narrowband CRIS scheme~\cite{Koszorus2020}, the atoms are excited from their initial hyperfine states by one or two consecutive resonant transitions. The spectroscopy transition is scanned in high resolution by changing the wavelength of a tunable narrowband laser. For transitions in the titanium sapphire (Ti:Sa) range, an injection-seeded cavity with a linewidth of $\sim$20~MHz is used~\cite{Sonnenschein2017}, while a pulsed dye amplifier with a typical linewidth of $\sim$150~MHz is used for the dye ranges. For resonant transitions excited in lower resolution, a broadband Ti:Sa or dye laser with a linewidth of ~$\sim$3-10~GHz is used. A high-power Nd:YAG laser is used for non-resonant ionization~\cite{Vernon2020}.

\section{Upgrades at CRIS} \label{sec:upgrades}
To optimize the performance of the CRIS experiment, technical upgrades to the beamline and laser systems are continuously developed, aiming to improve the spectroscopic background, efficiency, resolution, and precision of the technique. Previously, progress was reported on the development of a prototype field-ionization unit to reduce the background~\cite{Vernon2020}, the design of a radiofrequency cooler-buncher~\cite{Ricketts2020}, and the installation of $\alpha$-~\cite{Lynch2014a} and $\beta$-tagging~\cite{Koszorus2019thesis,Koszorus2021} stations to filter out stable and long-lived contaminants from the radioactive species. Recently, the development focus at the CRIS experiment has been placed on the implementation of voltage scanning and the upgrade of the beamline sections beyond the IR (referred to herein as \textit{the new end of the beamline}). 

\subsection{Voltage scanning} \label{subsec:voltage_scanning}
Instead of directly scanning the wavelength of the laser, CLS most often exploits the velocity of the fast beam and the Doppler effect to perform spectroscopy~\cite{Neugart2017}. The fast beam can be brought in and out of resonance with a fixed-wavelength laser in its rest frame by introducing slight variations to the beam's kinetic energy. This approach is a staple in fluorescence-detection CLS and is often referred to as \textit{voltage scanning}, since the kinetic energy is varied by applying varying voltages to a set of electrodes.

Voltage scanning offers significant benefits over laser scanning, and is thus strongly preferred in collinear experiments. By performing spectroscopy with voltage scanning, the narrowband laser system remains locked to a single wavelength and thus reaches greater stability over time, minimizing power fluctuations and leading to less noisy spectra. Additionally, the slew rate of the voltage-scanning electronics is typically considerably faster than the laser-scanning feedback loop based on the wavemeter reading, which measures the laser wavelength. As a result, for experiments on species with a sufficiently high production rate (in the order of $\sim$10$^{3}$ ions per second), voltage scanning can significantly accelerate the experimental run-time. 

However, there is a caveat in voltage scanning; accelerating an ion beam also electrostatically focuses it, while decelerating de-focuses it. As a result, in CLS beamlines employing voltage scanning, the detection region has so far been placed as close as possible after the voltage-scanning electrode. The CRIS beamline, where the IR is 1.2-m long and the detector is after an electrostatic bend (see Fig.~\ref{fig:New_end_beamline}), is particularly sensitive to changes in the ion-beam focus. For CLS and CRIS experiments on neutral atoms, the voltage-scanning electrode must be placed before the CEC, further increasing the distance to the detector. Previous attempts to implement voltage scanning at CRIS~\cite{Flanagan2013zakopane, Procter2013thesis} were thus unsuccessful as the ion transport through the beamline and onto the detector was significantly impacted by the changes in ion focus.

To bypass the past limitations related to ion transport, the new voltage-scanning setup at CRIS utilizes the electrode design reported by Gins et al. in Ref.~\cite{Gins2019}. The change in kinetic energy experienced by the ion beam while flying through the electrode is significantly smoother compared to the traditional design of resistively connected rings~\cite{Minamisono2013} that was also used in the past at CRIS. As a result, the aberrations in the beam collimation are presumed to be minimized, as shown below.

\begin{figure*}
    \centering
    \includegraphics[width=450pt]{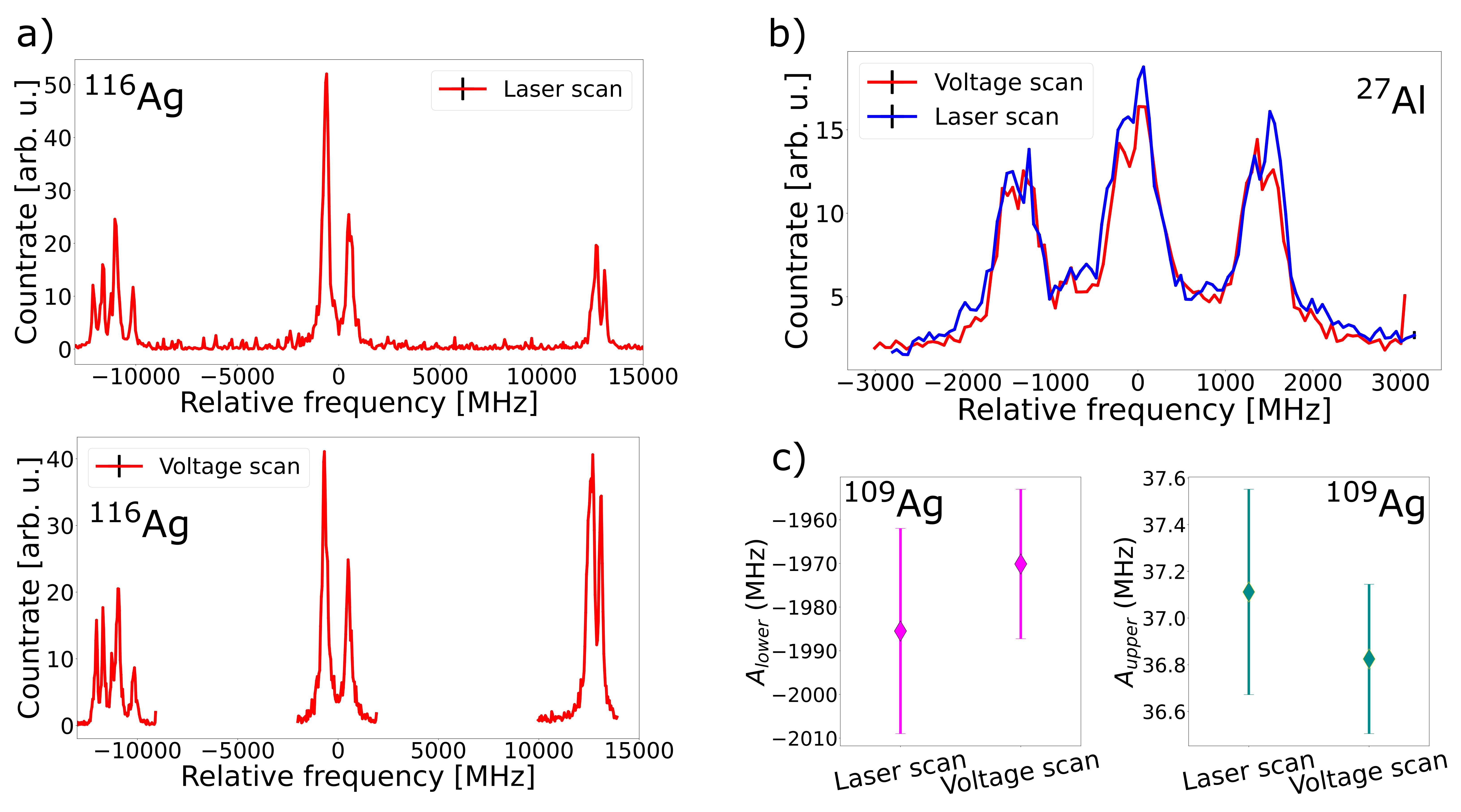}
    \caption{Commissioning tests of voltage scanning at CRIS using beams of $^{116}$Ag, $^{109}$Ag, and $^{27}$Al. \textbf{a)} Demonstration of background-skipping with voltage scanning. \textbf{b)} Visual comparison of spectra obtained with laser and voltage scanning, overlapped in the rest frame of the beam. \textbf{c)} Comparison of the hyperfine \textit{A} constants for the upper and lower electronic states in $^{109}$Ag. The results from the hyperfine fit for the voltage scans are fully consistent with those from the laser scans.}
    \label{fig:voltage_scanning}
\end{figure*}

The voltage-scanning setup was commissioned on beams of $^{109,116}$Ag and $^{27}$Al, comparing the results of frequency- and voltage-scanning. As seen in Fig.~\ref{fig:voltage_scanning}, the results of the hyperfine-structure fits are consistent for both scanning approaches. As no compromise in the spectroscopic efficiency was observed, it is concluded that the ion transport due to the changes in beam focus is largely unaffected. A voltage range of 400~V was successfully tested with no efficiency loss during the commissioning tests, which corresponds to more than 4~GHz in frequency space in the case of Ag. 

Voltage scanning led to a 3-6 times reduced time per scan in the cases of Ag and Al, depending on the isotope and its hyperfine structure. Overall, a potential time reduction per scan of up to 10 times is predicted from voltage scanning depending on the production yield, the wavelength range, the hyperfine structure, and the spectral resolution.

\subsection{New end of the beamline} \label{subsec:new_end}
\begin{figure*}
    \centering
    \includegraphics[width=480pt]{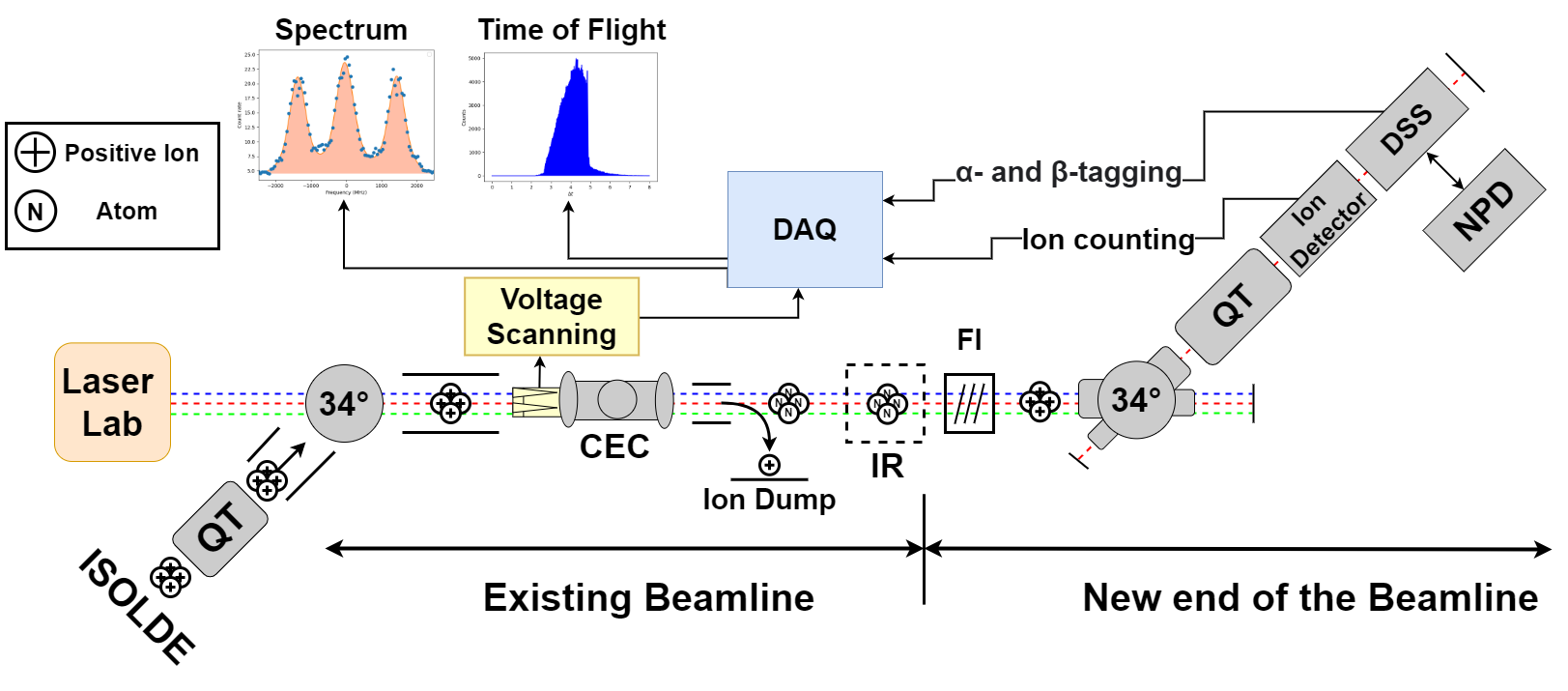}
    \caption{Schematic diagram of the existing CRIS beamline coupled to the new end of the beamline. The new end connects to the existing beamline at the laser-atom interaction region (IR), after neutralization in the charge-exchange cell (CEC). The IR is followed by the energy-selective field ionization (FI) unit, the 34$^{\circ}$ bending chamber, a quadrupole triplet (QT), a MagneToF single-ion detector, and the interchangeable decay spectroscopy station (DSS) and neutral particle detector (NPD).}    \label{fig:New_end_beamline}
\end{figure*}

For the CRIS technique to continue to push its limits, a new design for the end of the beamline is under construction. The new design will replace the current beamline from the end of the laser-atom IR onwards, as shown in Figure~\ref{fig:New_end_beamline}. The upgraded section consists of an energy-selective field ionization unit, a 34$^{\circ}$ bending chamber with collinear and anti-collinear laser access, a quadrupole triplet, deflection plates, a removable single-ion detector, and a decay spectroscopy station (DSS) that allows for $\beta$- and $\alpha$-tagging. The DSS chamber will also be interchangeable with a neutral particle detector for experiments on the laser photodetachment of negative ions.

A new energy-selective field ionization unit offers the opportunity to substitute the conventional non-resonant laser ionisation step with the electrostatic ionization of Rydberg states, while separating the field-ionized species from collisionally ionized background. The design of the apparatus was firstly presented by Vernon et al. in Ref.~\cite{Vernon2020}. As the atoms undergo field ionization, they are accelerated by hundreds of eV while the collisional ions remain at the initial beam energy. This small increase enables their angular separation in the 34$^\circ$ bend. The resulting difference in the ion trajectories downstream allows for the collisional ions to be filtered with the use of mechanical slits or by adjusting the position of the single-ion detector to align with the ion beam trajectory. Field ionization also removes the requirement of a high-power ionization laser that significantly contributes to the spectroscopic background through the non-resonant ionization of contaminant species. These improvements are critical to reduce the background towards performing background-free spectroscopy on beams with isobaric contamination that is a million times more intense than the isotope of interest, which is vital for the study of radioactive isotopes with low production yields~\cite{deGroote2017}.

The angle of the new bending chamber is 70$\%$ greater than that of the existing bend. In addition to the improved angular separation of field-ionized and background ions, the increased bending angle allows for charge-state selection, such as between +1 and +2 ions. This is particularly advantageous for campaigns where ionic spectroscopy is preferred, such as Group II elements.
Furthermore, the new bending chamber has been designed with two axes of laser access: one aligned with the laser-atom IR and one aligned with the chambers after the bend. The former is used for collinear (or anti-collinear) laser spectroscopy, whereas the latter can be used to perform photodetachment of negative ions in future campaigns.

Following the bending chamber, a quadrupole triplet and a set of deflector plates provide further control of the ion transport to the single-ion detector (MagneToF) and the DSS. The DSS will be equipped with $\alpha$ or $\beta$ detectors and a fast tape station to allow for decay-assisted laser spectroscopy~\cite{Lynch2014a} and laser-assisted decay spectroscopy, separating the isotopes depending on their decay mechanisms and energy.

\section{Summary and outlook} \label{sec:conclusion}
In conclusion, the CRIS experiment has been recently upgraded with the addition of a voltage-scanning apparatus to perform laser spectroscopy on short-lived species. Through voltage scanning, the time per scan of CRIS experiments can be potentially performed up to 10 times faster for isotopes produced with high yields (more than few thousand ions per second). 
Commissioning tests on $^{109,116}$Ag and $^{27}$Al have demonstrated that the hyperfine spectra measured with voltage and laser scanning are consistent. A voltage range of 400~V, which corresponds to a frequency range of $\sim$4~GHz in the case of Ag, was confirmed to cause no significant changes in the beam transport that would compromise the accuracy or efficiency of the CRIS experiment.

Additionally, upgrades to several regions of the CRIS beamline have been designed and are currently under construction and testing. The new regions include a field-ionization unit, a sharper electrostatic bend, and improved ion optics after the bend towards a single-ion counter and a decay spectroscopy station. Incorporating a field-ionization unit and a sharper bend leading to the ion detector are predicted to improve the signal-to-noise ratio by a factor of 5 or more, with the possibility of background-free spectroscopy when combined with the decay station.

\section*{Acknowledgments}
This project has received funding from FWO through IRI and regular projects, as well as from the Excellence of Science (EOS) programme (nr. 40007501), and the KU Leuven projects GOA15/010 and C14/22/104. The Department of Energy, Office of Science award DE-SC0021176 is also acknowledged.
 % \bibliographystyle{elsarticle-num}
 % \bibliography{references}

\end{document}